\documentstyle[amssymb,aps,prl]{revtex}
\begin{document}
\twocolumn[\hsize\textwidth\columnwidth\hsize\csname
@twocolumnfalse\endcsname

\draft{}
\title{CaB{\boldmath$_6$}: a new semiconducting material for spin 
electronics.}

\author{H.J. Tromp$^1$, P. van Gelderen$^{1,2}$, P.J. Kelly$^1$,
G. Brocks$^1$, and P.A. Bobbert$^3$.}

\address{
$^1$Faculty of Applied Physics and MESA+ Research Institute, 
University of Twente,
\\ P.O. Box 217, 7500 AE Enschede, The Netherlands\\
$^2$Faculty of Sciences and Research Institute for Materials, 
University of Nijmegen,\\ 
P.O. Box 9010, 6500 GL Nijmegen, The Netherlands\\
$^3$Department of Applied Physics and COBRA Research School, 
Eindhoven University of Technology,\\ 
P.O. Box 513, 5600 MB Eindhoven, The Netherlands}
\date{\today}
\maketitle
\begin{abstract}
Ferromagnetism was recently observed at unexpectedly high 
temperatures in La-doped CaB$_6$.
The starting point of all theoretical proposals to explain this
observation is a semimetallic electronic structure calculated for
CaB$_6$ within the local density approximation. Here we report the 
results of parameter-free quasiparticle calculations of the 
single-particle excitation spectrum which show that CaB$_6$ is not 
a semimetal but a semiconductor with a band gap of 0.8 eV. Magnetism 
in La$_x$Ca$_{1-x}$B$_6$ occurs just on the metallic side of
a Mott transition in the La-induced impurity band. 
\end{abstract}
\vskip2pc ]

\narrowtext

The recent observation of ferromagnetism in La doped alkaline-earth
hexaboride compounds at high temperatures{\cite{Young}} presents three 
puzzles.  Firstly, ferromagnetism is usually associated with elements 
with a partly filled 3d, 4f or 5f shell. Secondly, ferromagnetic 
ordering is only observed for a narrow dopant concentration range and 
for a surprisingly low dopant concentration. The maximum observed 
moment in Ca$_{1-x}$La$_x$B$_6$ is 0.07 Bohr magnetons per La atom for 
$x=0.005$. Thirdly, and most surprising of all is the observation of 
very large Curie temperatures; for La$_{0.01}$Ca$_{0.99}$B$_6$ a value 
close to 900 K has recently been reported{\cite{Ott}}. 
Such a large value virtually excludes the possibility of the 
observed magnetism being related to magnetic impurities.

In an initial comment on the experimental observations, Ceperley
suggested{\cite{Ceperley}} that it might be an example of the 
long-predicted but never observed ferromagnetic phase of a dilute 
electron gas; improved calculations increase the estimated density 
at which this might occur{\cite{Ortiz}}. An alternative explanation 
proposed by Zhitomirsky was that the ferromagnetic hexaborides might 
be doped excitonic insulators {\cite{Zhitomirsky}}. This explanation 
requires that the exciton binding energy should be comparable in 
size to the single-particle band gap. Jarlborg 
suggested{\cite{Jarlborg}} that the magnetism may be conventional 
itinerant magnetism. Although magnetism in materials which do not 
have partly filled d or f shells is rare, it is not unprecedented.
A handful of materials does exist, of which ZrZn$_2$ is the best
known example, in which the Fermi energy falls at or close to an
exceptionally narrow peak in the electronic density of states so
that the Stoner criterion for the occurence of itinerant magnetism
is fulfilled. The problem posed by the hexaborides is not so much 
the occurence of magnetic ordering but rather the strength of the 
magnetic coupling as reflected by the very high Curie temperature, 
and this was not estimated.

All three suggestions are based on electronic structure calculations
which indicate that stoichiometric CaB$_6$ is a semimetal with a 
very small overlap between the conduction and valence bands 
\cite{Hasegawa,Massidda,Rodriguez} or might have a very small gap
{\cite{Hasegawa}}. Since the theoretical predictions rely on details 
of this electronic structure, it is important to examine their validity.
Massidda's and Rodriguez' studies of the electronic structure were
based on the FLAPW calculation method which is probably the most
accurate available. These calculations were carried out within the 
framework of density functional theory (DFT) using the local density 
approximation (LDA). Such calculations are known to be capable of 
yielding total energies with a very useful accuracy.
However, it is also well known that the eigenvalue spectrum which
results from solving the Kohn-Sham equations of DFT cannot be
interpreted unreservedly as an excitation spectrum{\cite{Perdew,Sham}}
and sometimes the results are spectacularly wrong.
In particular, the band gaps of semiconductors are typically
underestimated by 50\% and in extreme cases such as Ge the 
conduction and valence bands are found to overlap resulting in 
metallic or semimetal character.

In order to study single-particle excitations, one should solve 
Dyson's equation for the single-particle Green's function expressed 
in terms of the self-energy operator $\Sigma$.
$\Sigma$ can be expanded as a perturbation series in the Green's
function $G$ and the dynamically screened Coulomb interaction $W$.
The so-called $GW$ approximation introduced by Hedin{\cite{Hedin}}
includes only the first term in this series. In addition, one 
usually assumes a quasi-particle (QP) expression for the Green's 
function $G$. For a large number of semiconductors and insulators 
such calculations produce band gaps which are in very close 
quantitative agreement with experimental single-particle band 
gaps{\cite{Aryasetiawan}}. The main purpose of the present 
paper is to apply these parameter-free calculations with their 
proven predictive capability to CaB$_6$.  From our results we shall 
conclude that, contrary to what is currently being assumed, the 
parent material is actually a semiconductor with quite a large 
band gap. This finding has far-reaching consequences for 
understanding the basic properties of the doped, ferromagnetic 
phase and opens up the prospect of a range of novel applications.

\begin{figure}
\begin{picture}(8.5,12.3)
\includegraphics{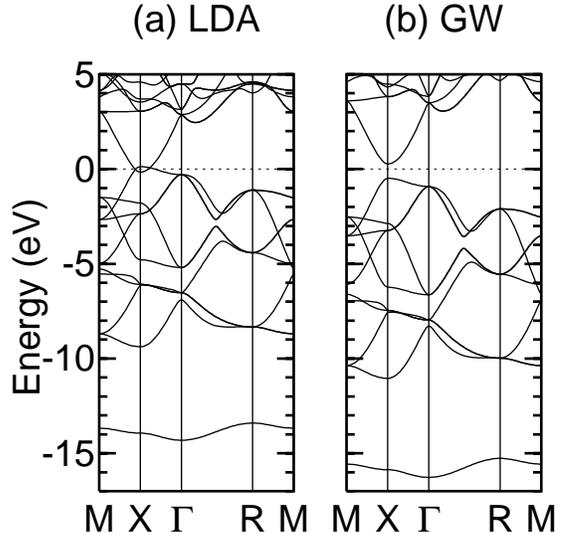}
\end{picture}
\vspace{7.5cm}

\caption{
Energy bands for CaB$_6$ in
(a) the LDA approximation,
(b) the $GW$ approximation.
The Fermi energy is indicated by the dotted line at 0 eV.}
\label{fig:all}
\end{figure}

The starting point for our $GW$ study is an LDA calculation for
CaB$_6$ performed with a plane-wave basis and using norm-conserving
pseudopotentials to describe the interaction between the valence
electrons and the ionic cores{\cite{Troullier}}. Using the 
experimentally determined structure, we reproduce essentially 
perfectly the energy bands calculated by Massidda{\cite{Massidda}}
using an all-electron method.
The results are shown in Fig.~\ref {fig:all}(a).
There are ten valence bands (bonding orbitals derived from the boron
2s and 2p levels) well-separated from the conduction bands
(antibonding states) with the exception of an overlap of about 0.3 
eV in a small region around the X point of the highest occupied and
lowest unoccupied bands. The occurrence of such small band overlap 
(or the presence of a comparably small band gap) is an important 
criterion for the stability
of an excitonic insulator phase{\cite{Zhitomirsky}}.
The effective masses in units of the free electron mass for the 
bands near X, $m_e^{\bot} = 0.22$ and $m_e^{\Vert} = 0.50$ for the
electrons and $m_h^{\bot} = 0.20$ and $m_h^{\Vert} = 1.8$ for 
the holes, are close to values reported earlier
{\cite{Hasegawa,Zhitomirsky}}.
We use the LDA electronic wave functions and eigenvalues as input 
for the $GW$ calculations for which we adopt the space-time approach
suggested by Rojas{\cite{Rojas}} in which all operators are represented
on grids in real and reciprocal space, in the time and energy domains;
details of our implementation have been given by van der 
Horst{\cite{van der Horst}}. By varying the size of these grids we 
estimate that the calculated QP energies are converged within 0.1 eV.
The densest grids we used were a $(12\times 12\times
12)$ real-space grid and a $(6\times 6\times 6)$ {\bf k}-grid. 
The Green's function is constructed including the LDA wave functions
and energies of the lowest 300 bands. 

The results of the $GW$ calculations are plotted in 
Fig.~\ref {fig:all}(b).
Throughout the Brillouin zone, the QP corrections to particular LDA
bands are rather uniform so that the dispersion of the $GW$ bands is
very similar to the LDA band dispersion and the effective masses 
differ only slightly from the values obtained from the LDA calculation.
They are reduced on average by less than 10\%, to $m_e^{\bot} = 0.20$ 
and $m_e^{\Vert} = 0.48$ for the electrons and $m_h^{\bot} = 0.17$ 
and $m_h^{\Vert} = 1.7$ for the holes. The sign and size of the 
corrections to the energy bands depend on their energy and their 
wavefunction character and vary considerably from band to band. 
Of particular importance are the relative shifts of the bands at X
near the Fermi level. The hole band and electron bands are moved, 
respectively, downwards and upwards in energy resulting in the 
opening of a sizeable band gap of about 0.8 eV. The relative shifts 
can be understood from the bonding and antibonding character of the 
wave functions in the valence and conduction bands, respectively. 
This is quite analogous to the situation in silicon{\cite{Godby}}.
The largest shift in the occupied bands is calculated for the lowest
valence band which is lowered in energy by about 1.9 eV.

Our finding that bulk CaB$_6$ is a semiconductor is consistent with 
the large value of the low-temperature resistivity and with its increase
with decreasing temperature recently reported by Ott{\cite{Ott}} and
by Vonlanthen\cite{Vonlanthen} and also with recent results of 
angle-resolved photoemission experiments{\cite{Denlinger}} in which 
a gap of a size similar to what we calculate was found.
Because of the sensitivity of photoemission experiments to surface
electronic structure, the authors interpreted their results 
exclusively in terms of the surface electronic structure.
Our result indicates that there might indeed be a slight increase in
the value of the band gap at the surface, something which is known 
to happen in heteropolar semiconductors.

The result for the stoichiometric phase has important implications 
for the models proposed to explain the ferromagnetism observed in the 
doped phase.  Using our calculated values for the effective masses and 
static dielectric constant of 5, we obtain a scaled density parameter 
$r_{\rm s}$ of 1.5, and an exciton binding energy of about 0.07 eV.
The scaled density parameter is still much lower than the values of
80 and 20 reported respectively, by Ceperley{\cite{Ceperley}} and
Ortiz{\cite{Ortiz}}, for which a ferromagnetic phase of the low 
density electron gas might be expected. The estimated excitonic 
binding energy is much smaller than the calculated QP band gap 
whereas the excitonic insulator models require
them to be comparable in size{\cite{Zhitomirsky}}.
Instead we note that the ferromagnetism occurs for an impurity
concentration which is just on the metallic side of a Mott
transition in the impurity band: $n^{1/3} a_H = 0.4$, where $n$ is
the dopant concentration $7 \times 10^{19}$ electrons cm$^{-3}$
and $a_H = 10$ \AA $\:$ is the Bohr radius of an isolated effective
mass impurity.

Support for this possibility can be found in the very recent
experimental studies on doped hexaborides reported by 
Terashima\cite{Terashima}.
He found that some of his La-doped samples were paramagnetic,
exhibiting the Curie-Weiss behaviour which is to be expected if the
dopant concentration were below Mott's critical value.
In other samples he found no indication of saturation of the magnetic
moment even in the highest magnetic field reported (10 kOe). 
This behaviour might be expected for partly filled narrow bands. 
In addition Terashima found much larger values of the magnetic 
moment per dopant atom than reported by Young and Ott\cite{Young,Ott}.
He also attributed a number of reported de Haas-van Alphen 
oscillation frequencies to Al inclusions.
We note that it is extremely important to know where the La goes and
if all of it actually occupies Ca sites as usually assumed since the
effective valence of an impurity atom in a semiconductor is very
sensitive to the local environment. 
Assuming that the magnetism occurs mainly in the impurity band, we 
are currently attempting to estimate the Curie temperature from 
total energy calculations and to determine the parameters which govern the
physical behaviour of the doped system such as the hopping matrix
elements, the charge and spin fluctuation parameters (Hubbard U
and Stoner I, respectively) and the binding energy of the impurity
state. The last quantity, the position of the impurity state with respect 
to the bottom of the conduction band, is particularly difficult to
estimate reliably. It requires knowing the central cell potential which 
is known to make a substantial correction to the binding energy of the 
lowest effective mass states.   

Irrespective of the origin of the high Curie temperature in the doped 
material, the existence of a large gap for the undoped material 
means a new class of magnetic semiconductors{\cite{Ohno1}} is available 
with the unique characteristic of having a Curie temperature sufficiently
high to allow room-temperature operation of semiconducting spin 
devices, something which has only very recently been achieved at low
temperatures{\cite{Fiederling,Ohno2}}. With the hexaborides it should be 
possible to inject a spin-polarized
current from the doped into the undoped material in the ballistic 
regime and study the spin dynamics as a function of temperature, and
of current density without the complication of having to apply an
external magnetic field{\cite{Fiederling,Ohno2}} or without the
problems encountered when attempting to inject spins from a
ferromagnetic metal into a conventional semiconductor in the 
diffusive regime{\cite{Schmidt}}. It should also be possible to make 
a field-effect device to modulate a spin-polarized current in the 
doped hexaboride material and achieve gain. If a $p$-type hexaboride 
with some of the Ca replaced by a monovalent
ion were also found to be ferromagnetic, it should even be possible
to make the spin-analogues of bipolar devices. 

Both Ca and B are light elements so that the spin-orbit coupling is
small.  Since the lowest conduction band and highest valence band are 
both singly degenerate, the intrinsic spin-flip scattering should be
weak with spin-flip scattering lengths comparable to what has 
recently been observed in carbon nanotubes{\cite{Tsukagoshi}}.
Because of the small size of the spin-orbit coupling, because the
magnetic moment is so tiny, and because the hexaborides are cubic,
the material should be magnetically extremely soft with a very small
intrinsic magnetocrystalline anisotropy.
This could be important for sensor applications.

P.J. Kelly is grateful to G.E.W. Bauer and J. van den Brink for 
useful discussions. P. van Gelderen and P.A. Bobbert wish to 
acknowledge discussions with R. Monnier and S. Massidda. This 
work is also part of the research program of the Stichting voor
Fundamenteel Onderzoek der Materie (FOM), financially supported by 
the Nederlandse Organisatie voor Wetenschappelijk Onderzoek (NWO) 
and Philips Research.

\end{document}